\documentclass[aps,nofootinbib,twocolumn,prl,superscriptaddress,amsmath,amssymb,amsfonts,floatfix]{revtex4-2}

\usepackage[dvips]{graphicx}
\usepackage{epsfig}
\usepackage{tabularx}
\usepackage{color}
\usepackage[colorlinks=true,linkcolor=blue,citecolor=magenta,urlcolor=blue]{hyperref}
\usepackage[mathscr]{eucal}
\usepackage{braket}
\usepackage{physics}

\usepackage{tikz,pgfplots}
\pgfplotsset{compat=1.18}
\usepackage{comment}
\usepackage{amsmath,amssymb,bm}
\usepackage{xcolor}
\usepackage{tikz}
\usepackage{pgfplots}
\pgfplotsset{compat=1.18}
\usetikzlibrary{arrows.meta,calc,positioning,decorations.pathmorphing,decorations.pathreplacing,patterns,3d}

\definecolor{cSample}{RGB}{200,0,0}     
\definecolor{cReg}{RGB}{0,92,175}       
\definecolor{cNonReg}{RGB}{230,110,30}  
\definecolor{cData}{RGB}{140,90,0}      

\RequirePackage[normalem]{ulem} 
\RequirePackage{color}\definecolor{RED}{rgb}{1,0,0}\definecolor{BLUE}{rgb}{0,0,1} 
\providecommand{\DIFaddbegin}{} 
\providecommand{\DIFaddend}{} 
\providecommand{\DIFdelbegin}{} 
\providecommand{\DIFdelend}{} 
\providecommand{\DIFaddbeginFL}{} 
\providecommand{\DIFaddendFL}{} 
\providecommand{\DIFdelbeginFL}{} 
\providecommand{\DIFdelendFL}{} 
\newcommand{\DIFscaledelfig}{0.5}
\RequirePackage{settobox} 
\RequirePackage{letltxmacro} 
\newsavebox{\DIFdelgraphicsbox} 
\newlength{\DIFdelgraphicswidth} 
\newlength{\DIFdelgraphicsheight} 
\LetLtxMacro{\DIFOincludegraphics}{\includegraphics} 
\newcommand{\DIFaddincludegraphics}[2][]{{\color{blue}\fbox{\DIFOincludegraphics[#1]{#2}}}} 
\newcommand{\DIFdelincludegraphics}[2][]{
\sbox{\DIFdelgraphicsbox}{\DIFOincludegraphics[#1]{#2}}
\settoboxwidth{\DIFdelgraphicswidth}{\DIFdelgraphicsbox} 
\settoboxtotalheight{\DIFdelgraphicsheight}{\DIFdelgraphicsbox} 
\scalebox{\DIFscaledelfig}{
\parbox[b]{\DIFdelgraphicswidth}{\usebox{\DIFdelgraphicsbox}\\[-\baselineskip] \rule{\DIFdelgraphicswidth}{0em}}\llap{\resizebox{\DIFdelgraphicswidth}{\DIFdelgraphicsheight}{
\setlength{\unitlength}{\DIFdelgraphicswidth}
\begin{picture}(1,1)
\thicklines\linethickness{2pt} 
{\color[rgb]{1,0,0}\put(0,0){\framebox(1,1){}}}
{\color[rgb]{1,0,0}\put(0,0){\line( 1,1){1}}}
{\color[rgb]{1,0,0}\put(0,1){\line(1,-1){1}}}
\end{picture}
}\hspace*{3pt}}} 
} 
\LetLtxMacro{\DIFOaddbegin}{\DIFaddbegin} 
\LetLtxMacro{\DIFOaddend}{\DIFaddend} 
\LetLtxMacro{\DIFOdelbegin}{\DIFdelbegin} 
\LetLtxMacro{\DIFOdelend}{\DIFdelend} 
\DeclareRobustCommand{\DIFaddbegin}{\DIFOaddbegin \let\includegraphics\DIFaddincludegraphics} 
\DeclareRobustCommand{\DIFaddend}{\DIFOaddend \let\includegraphics\DIFOincludegraphics} 
\DeclareRobustCommand{\DIFdelbegin}{\DIFOdelbegin \let\includegraphics\DIFdelincludegraphics} 
\DeclareRobustCommand{\DIFdelend}{\DIFOaddend \let\includegraphics\DIFOincludegraphics} 
\LetLtxMacro{\DIFOaddbeginFL}{\DIFaddbeginFL} 
\LetLtxMacro{\DIFOaddendFL}{\DIFaddendFL} 
\LetLtxMacro{\DIFOdelbeginFL}{\DIFdelbeginFL} 
\LetLtxMacro{\DIFOdelendFL}{\DIFdelendFL} 
\DeclareRobustCommand{\DIFaddbeginFL}{\DIFOaddbeginFL \let\includegraphics\DIFaddincludegraphics} 
\DeclareRobustCommand{\DIFaddendFL}{\DIFOaddendFL \let\includegraphics\DIFOincludegraphics} 
\DeclareRobustCommand{\DIFdelbeginFL}{\DIFOdelbeginFL \let\includegraphics\DIFdelincludegraphics} 
\DeclareRobustCommand{\DIFdelendFL}{\DIFOaddendFL \let\includegraphics\DIFOincludegraphics} 

\begin{document}

\title{Finite-Resolution Limits on Quantum Uniqueness}
\DIFdelbegin 
\author{Seramika Ariwahjoedi}
\DIFaddbegin \email{ceramyca@gmail.com}
\DIFaddend \affiliation{Research Center for Quantum Physics, National Research and Innovation Agency (BRIN), South Tangerang 15314, Banten, Indonesia  
}
\DIFaddend 
\DIFdelend \author{Sameer Ahmad Mir}
\DIFdelbegin 
\DIFdelend \DIFaddbegin \email{sameerphst@gmail.com}
\DIFaddend \affiliation{Canadian Quantum Research Center, 204-3002 32 Ave, Vernon, BC V1T 2L7, Canada}
\affiliation{Department of Computer Sciences, Asian School of Business, Noida, Uttar Pradesh, 201303, India}
\DIFaddend

\author{Mir Faizal}
\DIFaddbegin \email{mirfaizalmir@googlemail.com}
\DIFaddend \affiliation{Canadian Quantum Research Center, 204-3002 32 Ave, Vernon, BC V1T 2L7, Canada}
\affiliation{Irving K. Barber School of Arts and Sciences, 
  University of British Columbia Okanagan, Kelowna,
British Columbia V1V 1V7, Canada}
\affiliation{Department of Mathematical Sciences, Durham University, Upper Mountjoy, Stockton Road, Durham DH1 3LE, UK}
\affiliation{Faculty of Sciences, Hasselt University, Agoralaan Gebouw D, Diepenbeek, 3590, Belgium}

\DIFdelbegin 
\DIFdelend 

\begin{abstract}
The Stone-von Neumann theorem makes canonical quantization unique only after one imposes strong continuity of the Weyl translation groups. We show that this continuity assumption cannot be certified by any finite-resolution interrogation of the Weyl relations. A finite protocol probes only finitely many phase-space displacements and bounded expectation values, thereby determining a protocol algebra \(C_S\) generated by a finitely generated subgroup of Weyl translations, but not the regularity of an extension to the full Weyl algebra. Using the polymer representation as a canonical non-regular representation, we prove that for every regular Weyl representation, every normal state, and every finite family of bounded protocol observables, a polymer state reproduces the same expectation values to arbitrary prescribed accuracy. Thus finite agreement with Schr\"{o}dinger quantum mechanics does not operationally certify the regularity hypothesis entering uniqueness. The result does not assert physical equivalence between regular and polymer quantizations, it identifies where additional input, such as Hamiltonian dynamics, energy regularity, semiclassicality, or a continuum limit, is required to distinguish or select representations. In this precise sense, non-regular polymer kinematics are not excluded by finite Weyl data alone, a point of direct relevance to quantum-gravity motivated quantizations.
\end{abstract}

\maketitle

\paragraph{Introduction.}

Canonical quantization of a single degree of freedom is most naturally formulated in terms of Weyl operators implementing phase-space translations and satisfying the Weyl form of the canonical commutation relations \cite{Weyl}. The standard uniqueness result in this setting is the Stone-von Neumann theorem, which asserts that all irreducible representations of the Weyl relations are unitarily equivalent provided the translation groups are strongly continuous \cite{Stone,vonNeumann}. Strong continuity is the sole analytic hypothesis of the theorem; it permits the use of Stone's theorem to obtain self-adjoint generators and hence the familiar position and momentum operators \cite{ReedSimon,Dixmier}. While mathematically natural, strong continuity is not an operational requirement implied by finite experiments. Laboratory protocols implement translations through finitely many calibrated control settings with finite resolution and finite statistics, so only a finite family of Weyl operators and their finite compositions can be prepared and distinguished, up to bounded tolerances in their parameters. The Stone-von Neumann theorem does not address whether strong continuity can be certified, or excluded, from such finite operational data.

In this Letter we formulate finite experimental resolution directly at the level of accessible Weyl operators, without modifying the Weyl relations and without privileging any representation a priori. The point is to separate two logically distinct ingredients of the usual uniqueness statement. The Weyl relations fix the algebraic composition law of phase-space translations, whereas regularity is an additional strong-continuity condition imposed on a representation of that algebra. This distinction belongs naturally to the \(C^*\)-algebraic formulation of canonical commutation relations and unitary group representations \cite{BratteliRobinson,FellDoran,KadisonRingrose,Segal}. A finite experiment probes only finitely many displacement settings and finitely many expectation values. It therefore determines only a finite operational algebra \(C_S\), not the strong-continuity behavior of the full one-parameter Weyl translation groups.

Our main result shows that this limitation is structural, not merely practical. For any finite protocol algebra \(C_S\), any regular Weyl representation, any normal state, and any finite family of bounded observables in \(C_S\), the same expectation values can be reproduced to arbitrary prescribed accuracy. This reproduction is achieved by a suitable state in the polymer representation. The polymer representation is non-regular. The approximation mechanism uses only standard \(C^*\)-algebraic tools. The protocol algebra is a twisted group \(C^*\)-algebra of the finitely generated abelian group \(\Gamma_S\). Amenability identifies the relevant full and reduced twisted group \(C^*\)-algebras. Fell approximation then allows states to be approximated on finite sets by vector states in a faithful representation \cite{Hulanicki,Segal,Rieffel}. The polymer representation is used here only as a canonical non-regular realization of the Weyl relations. The representation-theoretic approximation argument relies on the standard algebraic facts described above \cite{Segal,FellDoran,Rieffel}. This does not contradict the Stone-von Neumann theorem. That theorem remains exactly valid within the class of irreducible regular representations. Our result identifies instead the operational status of its continuity hypothesis. Regularity is an additional analytic input. It is not something certified by finite access to Weyl translations alone.

\paragraph{Experimental relevance and finite operational access.}
Let \(\xi=(\lambda,\mu)\in\mathbb R^2\) denote a phase-space displacement
label, and let \(W(\xi)\) denote the corresponding Weyl displacement
unitary. In experiments, such phase-space displacements are implemented by
calibrated control pulses with finite resolution and are inferred from
finitely many measurement shots. The accessible information is therefore not
the full continuum of Weyl labels \(\xi\in\mathbb R^2\), but a finite data
table consisting of discrete control settings, uncertain displacement
parameters, and finitely sampled expectation values. Thus a finite protocol
can constrain only finitely many bounded operational quantities, each within
finite statistical and calibration tolerances.

In continuous-variable systems this restriction is explicit. Displacements,
realized as Gaussian primitives, are characterized indirectly through
rotated-quadrature measurements, homodyne sampling, and tomographic
reconstructions employing finitely many local-oscillator phases and finitely
many samples. These procedures constrain only a finite family of accessible
displacement operators, or coarse-grained approximants to them, together with
finite moment or correlation data
\cite{VogelRisken,Smithey,LvovskyRaymer,Weedbrook}. More generally,
validation of implemented dynamics relies on process tomography and
self-consistent variants, yielding only finite input-output probabilities
and a finite reconstructed model
\cite{ChuangNielsen,Merkel:2012tyt,Blume-Kohout:2017rik}, while randomized
benchmarking provides finite-data estimates of average error rates without
pointwise continuity information \cite{Knill:2007aqr,Magesan:2010tmc}.

This operational finiteness motivates our formulation. A protocol specifies
a finite set of calibrated displacement labels
\begin{equation}
        S=\{\xi_1,\ldots,\xi_m\}\subset\mathbb R^2 ,
\end{equation}
and finitely many expectation values estimated within finite tolerance. The
subgroup algebraically generated by the accessible labels is
\(
        \Gamma_S=\operatorname{span}_{\mathbb Z}(S)\subset\mathbb R^2 .
\)
The physically available control operations generate the unital
\(\ast\)-algebra spanned by finite products of
\(\{W(\xi_j)\}_{j=1}^m\). Equivalently, the relevant Weyl labels lie in
\(\Gamma_S\). The corresponding norm-closed protocol algebra is
\begin{equation}
        \mathcal C_S
        =
        C^\ast\big(\{W(\gamma):\gamma\in\Gamma_S\}\big)
        \subset \mathcal W ,
\label{eq:protocol-algebra}
\end{equation}
where \(\mathcal W\) denotes the full Weyl \(C^\ast\)-algebra. Thus
\(\mathcal C_S\) is the part of the Weyl algebra generated by the control
operations available to the protocol. The distinction is important:
\(\Gamma_S\), and hence \(\mathcal C_S\), may be infinite even though \(S\)
is finite. 
However, any actual finite data set tests only finitely many
bounded observables in \(\mathcal C_S\), and only up to finite tolerance. 
The notation \(\xi_n\to0\) refers to convergence in the Euclidean topology
of the full parameter space \(\mathbb R^2\). It should not be confused with
the experimental act of enlarging the finite set \(S\). Strong continuity is
a representation-theoretic condition: for a representation \(\pi\), it
requires
\begin{equation}
\begin{aligned}
        \lim_{\lambda\to0}
        \|\pi(W(\lambda,0))\psi-\psi\|=0,
        \\
        \lim_{\mu\to0}
        \|\pi(W(0,\mu))\psi-\psi\|=0
\end{aligned}
\label{eq.3}
\end{equation}
for every vector \(\psi\) in the representation Hilbert space. Equivalently,
since the Weyl parameter space is metrizable, regularity requires the same
convergence along every sequence \(\xi_n\to0\) in the Euclidean topology of
\(\mathbb R^2\). This is an infinite limiting condition at the identity of
the displacement group. A finite protocol supplies no such limiting test. It
samples only the displacement labels generated by finitely many chosen
settings and compares only finitely many bounded expectation values.

\begin{figure}[t]
\centering

\begin{minipage}[t]{0.48\textwidth}
\centering
\begin{tikzpicture}
\begin{axis}[
    width=\linewidth,
    height=0.90\linewidth,
    xmin=-2.5, xmax=2.5,
    ymin=-2.85, ymax=2.85,
    axis lines=box,
    axis equal image,
    xlabel={$q$},
    ylabel={$p$},
    xtick={-2,-1,0,1,2},
    ytick={-2,-1,0,1,2},
    grid=both,
    grid style={gray!20},
    tick label style={font=\small},
    label style={font=\small},
    clip=false,
]
\node[anchor=north west, font=\bfseries] at (rel axis cs:0.02,0.98) {(a)};

\addplot+[
    only marks,
    mark=*,
    mark size=6pt,
    draw=cSample,
    fill=cSample,
    fill opacity=0.12,
    draw opacity=0.35,
] coordinates {
    (-2,-2) (-2,-1) (-2,0) (-2,1) (-2,2)
    (-1,-2) (-1,-1) (-1,0) (-1,1) (-1,2)
    (0,-2)  (0,-1)  (0,0)  (0,1)  (0,2)
    (1,-2)  (1,-1)  (1,0)  (1,1)  (1,2)
    (2,-2)  (2,-1)  (2,0)  (2,1)  (2,2)
};

\addplot+[
    only marks,
    mark=*,
    mark size=2pt,
    mark options={fill=cSample, draw=cSample},
] coordinates {
    (-2,-2) (-2,-1) (-2,0) (-2,1) (-2,2)
    (-1,-2) (-1,-1) (-1,0) (-1,1) (-1,2)
    (0,-2)  (0,-1)  (0,0)  (0,1)  (0,2)
    (1,-2)  (1,-1)  (1,0)  (1,1)  (1,2)
    (2,-2)  (2,-1)  (2,0)  (2,1)  (2,2)
};

\node[anchor=center, font=\scriptsize] at (axis cs:0,3.05)
    {Finite protocol: Sampled settings $\{\xi_k\}$};

\draw[cSample, thick, ->] (axis cs:1.35,2.25) -- (axis cs:1.10,1.30);
\node[anchor=west, font=\scriptsize, text=cSample] at (axis cs:0.88,2.35)
    {tolerance $\delta\xi$};

\end{axis}
\end{tikzpicture}
\end{minipage}
\hfill

\begin{minipage}[t]{0.48\textwidth}
\centering
\begin{tikzpicture}
\begin{axis}[
    width=\linewidth,
    height=0.90\linewidth,
    xmin=-2.5, xmax=2.5,
    ymin=0.0, ymax=1.15,
    axis lines=left,
    xtick={-2,-1,0,1,2},
    ytick={0,0.5,1.0},
    xlabel={$\xi$},
    ylabel={$f(\xi)$},
    tick label style={font=\small},
    label style={font=\small},
    legend style={at={(0.98,1.00)}, anchor=north east, draw=none, fill=none, font=\small},
]
\node[anchor=north west, font=\bfseries] at (rel axis cs:0.02,0.98) {(b)};

\addplot[cReg, very thick, domain=-2.5:2.5, samples=220] {exp(-x^2/2)};
\addlegendentry{regular}

\addplot[cNonReg, thick, dashed] coordinates {
    (-2.5,0.10)
    (-2,0.1353)
    (-1.7,0.1353)
    (-1.7,0.95)
    (-1.3,0.95)
    (-1.3,0.6065)
    (-1,0.6065)
    (-0.8,0.6065)
    (-0.8,0.20)
    (-0.4,0.20)
    (-0.4,1.00)
    (0,1.00)
    (0.2,1.00)
    (0.2,0.30)
    (0.7,0.30)
    (0.7,0.85)
    (0.85,0.85)
    (0.85,0.6065)
    (1,0.6065)
    (1.3,0.6065)
    (1.3,0.95)
    (1.6,0.95)
    (1.6,0.1353)
    (2,0.1353)
    (2.5,0.20)
};
\addlegendentry{non-regular}

\addplot+[
    cData, only marks,
    mark=*,
    mark size=2.2pt,
    mark options={fill=cData, draw=cData},
    error bars/.cd,
        x dir=both, x explicit,
        y dir=both, y explicit,
        error bar style={cData},
] coordinates {
    (-2,0.1353) +- (0.18,0.06)
    (-1,0.6065) +- (0.18,0.06)
    (0,1.0000)  +- (0.18,0.06)
    (1,0.6065)  +- (0.18,0.06)
    (2,0.1353)  +- (0.18,0.06)
};

\end{axis}
\end{tikzpicture}
\end{minipage}

\caption{Finite-resolution access to Weyl translations (schematic). (a) Finite control precision restricts experimental access to a discrete family of displacement parameters $\{\xi_k\}\subset\mathbb{R}^2$ with $\xi=(q,p)$, each realized only within a tolerance $\delta\xi$ (disks). (b) Finite statistics determine only noisy estimates of an operational quantity, e.g. $f(\xi)=\langle\psi|W(\xi)|\psi\rangle$, at sampled settings; a regular (strongly continuous) and a non-regular realization can agree on all sampled points within error bars, so strong continuity is not operationally decidable from finite data.}
\label{fig:finite_resolution_weyl_sampling_qp}
\end{figure}
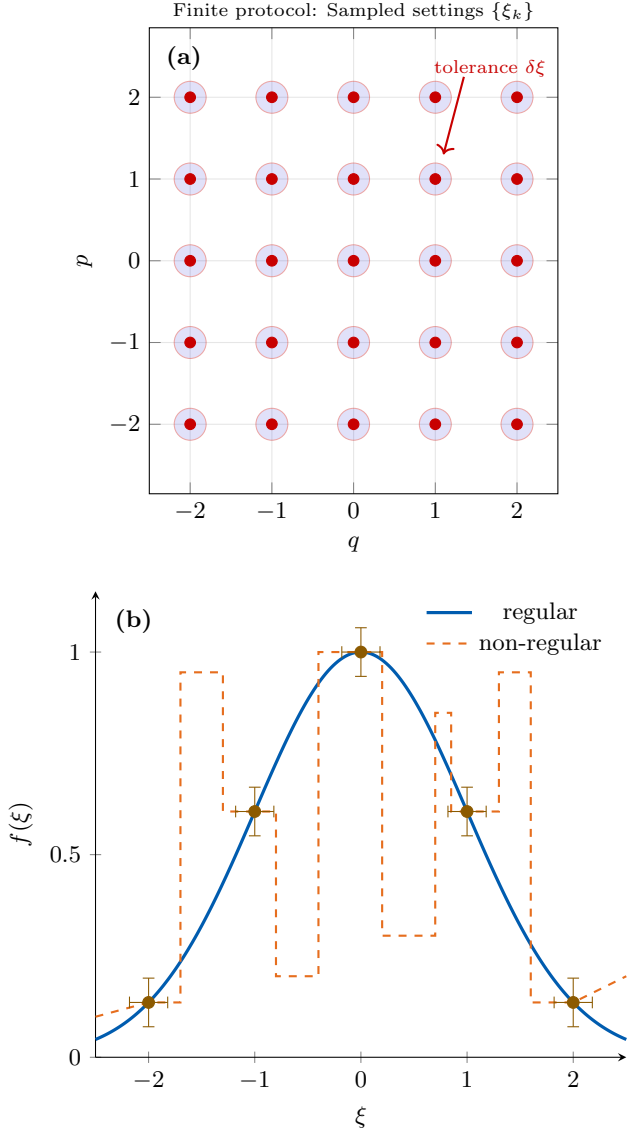

Allowing adaptive refinement does not alter the logical point. If a later
experiment uses additional settings, \(S\) is replaced by a larger finite set
and \(\mathcal C_S\) by a larger finitely generated protocol algebra. Each
finite stage may provide a better operational approximation to a continuum
model, but no finite stage certifies strong continuity of the full
one-parameter Weyl translation groups
\begin{equation}
        \lambda\mapsto \pi(W(\lambda,0)),
        \qquad
        \mu\mapsto \pi(W(0,\mu)).
\label{eq.4}
\end{equation}
Accordingly, \(\mathcal C_S\) encodes the discrete access structure of the
protocol, whereas regularity concerns an extension of a representation from
\(\mathcal C_S\) to the full Weyl algebra together with the Euclidean
topology of its label space. Regularity may therefore be a natural and
physically useful modeling assumption, but it is not forced by finite
operational access to Weyl translations. 
This does not mean that regular and non-regular representations are
physically identical in all respects. It means only that finite
Weyl-displacement data, by themselves, do not decide the strong-continuity
hypothesis required by the Stone--von Neumann uniqueness theorem. Additional
physical structure, such as Hamiltonian dynamics, energy regularity,
semiclassicality, or a continuum limit, may distinguish or select
representations. Such structure, however, goes beyond the finite protocol
algebra \(\mathcal C_S\). 
This finite-resolution limitation is summarized schematically in
Fig.~\ref{fig:finite_resolution_weyl_sampling_qp}. Finite control precision
selects only sampled displacement labels with error bars, and finite
statistics determine only finitely many noisy expectation values. A regular
representation and a non-regular representation may agree on all such sampled
data within the prescribed tolerance, even though they differ in their
behavior under the infinite limiting test required for strong continuity.

\paragraph{Weyl algebra and regularity.}
Let \(\xi=(\lambda,\mu)\) and \(\eta=(\lambda',\mu')\) be elements of
\(\mathbb R^2\). We use the standard symplectic form
\begin{equation}
        \sigma(\xi,\eta)=\lambda\mu'-\mu\lambda' .
\end{equation}
The Weyl \(C^*\)-algebra \(\mathcal W\) is the universal \(C^*\)-algebra
generated by unitaries \(\{W(\xi)\}_{\xi\in\mathbb R^2}\) satisfying
\begin{equation}
        W(\xi)W(\eta)
        =
        e^{-\frac{i}{2}\sigma(\xi,\eta)}W(\xi+\eta),
        \,\,
        W(\xi)^*=W(-\xi).
\label{eq:weyl}
\end{equation}
These relations are the Weyl form of the canonical commutation relations \cite{Weyl}. 
A representation of \(\mathcal W\) is a nondegenerate
\(*\)-homomorphism
\(
        \pi:\mathcal W\longrightarrow B(\mathcal H),
\)
where \(B(\mathcal H)\) denotes the bounded operators on the representation
Hilbert space \(\mathcal H\). 
Such a representation is called regular if the
one-parameter unitary groups as in Eq.~\eqref{eq.4}, are strongly continuous. 
Equivalently, regularity is strong continuity at the identity in the sense
stated in Eq.~\eqref{eq.3}: for every \(\psi\in\mathcal H\), the vectors
\(\pi(W(\lambda,0))\psi\) and \(\pi(W(0,\mu))\psi\) converge in
Hilbert-space norm to \(\psi\) as \(\lambda\to0\) and \(\mu\to0\),
respectively. 
Thus regularity is not an algebraic consequence of the Weyl relations alone.
It is an additional continuity requirement on the representation. 
Under this requirement, Stone's theorem gives self-adjoint generators \(Q\) and \(P\) for the two one-parameter unitary groups. The Stone-von Neumann theorem then states that every irreducible regular representation of the Weyl relations for one canonical degree of freedom  is unitarily equivalent to the Schr\"{o}dinger representation \cite{Stone,vonNeumann,ReedSimon,Dixmier}.

Therefore the theorem proves uniqueness inside the regular sector. 
It does not prove that finite operational data, restricted to a protocol
algebra \(C_S\), force a representation of the full Weyl algebra
\(\mathcal W\) to be regular.

\paragraph{Operational resolution.}
With the notation above, a finite protocol 
specified by \(S=\{\xi_1,\ldots,\xi_m\}\) determines the algebraically
generated subgroup
\(
        \Gamma_S=\operatorname{span}_{\mathbb Z}(S),
\)
and hence, through Eq.~\eqref{eq:protocol-algebra}, the associated protocol
algebra \(C_S\). 
Although \(C_S\) may contain infinitely many algebraic and norm-limit
elements, any actual finite experimental data set tests only finitely many
bounded observables in \(C_S\), and only within finite tolerance. Thus the
operational comparison between two representations is necessarily a
finite-set comparison of expectation values.

Let
\begin{equation}
        \pi_1:\mathcal W\to B(\mathcal H_1),
        \quad
        \pi_2:\mathcal W\to B(\mathcal H_2)
\end{equation}
be two representations. 
We say that \(\pi_2\) is \((S,\eta)\)-operationally compatible with \(\pi_1\) if, for every density operator \(\rho_1\) on \(\mathcal H_1\) and every finite collection of  test observables \(A_1,\ldots,A_n\in C_S\) with \(\|A_j\|\le1\), there exists a density operator \(\rho_2\) on \(\mathcal H_2\) such that
\begin{equation}
        \left|
        \operatorname{Tr}\big(\rho_1\pi_1(A_j)\big)
        -
        \operatorname{Tr}\big(\rho_2\pi_2(A_j)\big)
        \right|
        \le \eta ,
        \,\, j=1,\ldots,n .
\end{equation}
This is intentionally a one-sided notion. It formalizes the statement that
\(\pi_2\) cannot be ruled out by any finite \((C_S,\eta)\) data set
compatible with \(\pi_1\). In the application below, \(\pi_1\) will be a
regular Weyl representation and \(\pi_2\) will be the polymer representation.

The notation \(\xi_n\to0\) always refers to convergence in the Euclidean
topology of the full parameter space \(\mathbb R^2\). This is precisely the
topology used in the definition of regularity. A finite protocol does not
implement an arbitrary sequence approaching the identity
in this topology.
It accesses only the subgroup \(\Gamma_S\) generated by finitely many chosen settings, and its finite data constrain only finitely many expectation values on \(C_S\). Hence \(C_S\) records the algebraic information available to the protocol, whereas regularity concerns an extension of a representation from \(C_S\) to the full Weyl algebra \(\mathcal W\) together with the Euclidean topology of its label space. 
Therefore the Stone-von Neumann theorem yields uniqueness only after the
additional continuity assumption of regularity has been imposed. Finite
operational data on \(C_S\) may be consistent with a regular representation,
but they do not by themselves certify that every compatible extension to the
full Weyl algebra is regular.

\paragraph{Operational continuity and the scope of uniqueness.}
For a fixed protocol $S$, the accessible parameter set is $\Gamma_S$: only finitely many generators and their finite products are queried, and no operation internal to the protocol implements a limiting procedure $\xi_n\to 0$ in $\mathbb{R}^2$. Accordingly, the protocol induces no topology on $\Gamma_S$ beyond the discrete one, and in particular does not operationally test strong continuity of $\lambda\mapsto \pi(W(\lambda,0))$ or $\mu\mapsto \pi(W(0,\mu))$ at the identity. Regularity is therefore not a property of the abstract operational algebra $\mathcal{C}_S$, but of an extension of a representation of $\mathcal{C}_S$ to a representation of $\mathcal{W}$ together with the Euclidean topology on $\mathbb{R}^2$. The Stone-von Neumann theorem thus yields uniqueness only within the class of irreducible regular representations, while leaving regularity itself undetermined by operational data restricted to $\mathcal{C}_S$.

\paragraph{Polymer representation.}
Let \(\mathbb R_d\) denote the real line equipped with the discrete topology,
and define
\begin{equation}
        \mathcal H_{\rm pol}=\ell^2(\mathbb R_d).
\end{equation}
Here \(\mathbb R_d\) has the same underlying set as \(\mathbb R\), but it is
regarded as a discrete measure space with counting measure.
Thus \(\mathcal H_{\rm pol}\) is not the Schrödinger Hilbert space
\(L^2(\mathbb R)\). Its vectors are square-summable functions on the set
\(\mathbb R\) with respect to counting measure. Consequently the vectors
\(\{|x\rangle\}_{x\in\mathbb R}\) are genuine normalizable basis vectors of
\(\mathcal H_{\rm pol}\), satisfying
\(
        \langle x|x'\rangle=\delta_{x,x'} ,
\)
where \(\delta_{x,x'}\) is the Kronecker delta. This is different from the
usual generalized position eigenvectors of \(L^2(\mathbb R)\), which are
distributions and only formally satisfy Dirac-delta normalization. 
This distinction is essential: in \(\mathcal H_{\rm pol}\), two different
basis vectors \(|x\rangle\) and \(|x'\rangle\) are orthogonal whenever
\(x\neq x'\).

The polymer action of the Weyl generators is given by
\begin{equation}
        \pi_{\rm pol}(W(\lambda,\mu))|x\rangle
        =
        e^{i\mu(x+\lambda/2)}|x+\lambda\rangle .
\end{equation}
Each \(\pi_{\rm pol}(W(\lambda,\mu))\) is unitary, since it shifts the
orthonormal basis by \(x\mapsto x+\lambda\) and multiplies each basis vector
by a phase.
A direct evaluation on basis vectors gives
\begin{equation}
        \pi_{\rm pol}(W(\xi))\pi_{\rm pol}(W(\eta))
        =
        e^{-\frac{i}{2}\sigma(\xi,\eta)}
        \pi_{\rm pol}(W(\xi+\eta)).
\end{equation}
Thus the polymer construction is a genuine representation of the Weyl
relations. It is nevertheless non-regular \cite{Segal,Rieffel}. 
Indeed, for any \(x\in\mathbb R\) and any nonzero \(\lambda\),
the vectors \(|x+\lambda\rangle\) and \(|x\rangle\) are orthogonal in
\(\mathcal H_{\rm pol}\). Hence
\begin{equation}
        \big\|
        \pi_{\rm pol}(W(\lambda,0))|x\rangle-|x\rangle
        \big\|^2
        =
        \big\||x+\lambda\rangle-|x\rangle\big\|^2
        =
        2 .
\end{equation}
Therefore \(\lambda\mapsto\pi_{\rm pol}(W(\lambda,0))\) is discontinuous at \(\lambda=0\) in the strong operator topology. 
The failure of continuity is not due to a failure of the Weyl relations; rather, it
follows from the discrete Hilbert-space topology of the polymer
representation.

For the fixed protocol \(S\), the shift subgroup is defined as
\begin{equation}
        \Lambda_S
        =
        \{\lambda\in\mathbb R:(\lambda,\mu)\in\Gamma_S
        \text{ for some }\mu\}.
\label{eq:LambdaS}
\end{equation}
For each \(x_0\in\mathbb R\), we define
\begin{equation}
        \mathcal H_{x_0,S}
        =
        \overline{\operatorname{span}}
        \{|x_0+\lambda\rangle:\lambda\in\Lambda_S\}
        \subset\mathcal H_{\rm pol}.
\label{eq:polymer-sector}
\end{equation}
Each \(\mathcal H_{x_0,S}\) is invariant under
\(\pi_{\rm pol}(C_S)\), because every element of \(C_S\) shifts a basis
vector only by an element of \(\Lambda_S\). 
More precisely, the Weyl generators with labels in \(\Gamma_S\) shift basis
vectors only by elements of \(\Lambda_S\), and the invariance then extends
from finite algebraic combinations to the norm-closed algebra \(C_S\) because
\(\mathcal H_{x_0,S}\) is closed. 
Explicitly, for
\((\lambda,\mu)\in\Gamma_S\),
\begin{multline}
        \pi_{\rm pol}(W(\lambda,\mu))|x_0+\lambda'\rangle
        =
        e^{i\mu(x_0+\lambda'+\lambda/2)}
        |x_0+\lambda'+\lambda\rangle ,
        \\ \lambda'\in\Lambda_S .
\end{multline}
Since \(\lambda\in\Lambda_S\), the vector on the right again belongs to
\(\mathcal H_{x_0,S}\). 
Thus \(\mathcal H_{\rm pol}\) decomposes as an orthogonal direct sum of
invariant protocol sectors indexed by the cosets of \(\Lambda_S\) in
\(\mathbb R\). The proof below uses only this sector decomposition and
the standard finite-set form of Fell approximation for faithful
representations of \(C^*\)-algebras.

\paragraph{Main theorem.}
Fix a finite protocol \(S\), let
\(
        \Gamma_S=\operatorname{span}_{\mathbb Z}(S),
        \,\,
        C_S=C^*\big(\{W(\gamma):\gamma\in\Gamma_S\}\big),
\)
and let \(\eta>0\). Let
\(
        \pi:\mathcal W\longrightarrow B(\mathcal H)
\)
be a regular Weyl representation, and let \(\rho\) be a density operator on
\(\mathcal H\). For every finite family
\(A_1,\ldots,A_n\in C_S\) with \(\|A_j\|\le1\), there exists a density
operator \(\rho_{\rm pol}\) on \(\mathcal H_{\rm pol}\), supported on a
finite direct sum of invariant polymer sectors, such that
\begin{multline}
        \left|
        \operatorname{Tr}\big(\rho\,\pi(A_j)\big)
        -
        \operatorname{Tr}\big(\rho_{\rm pol}\,
        \pi_{\rm pol}(A_j)\big)
        \right|
        \le \eta ,
        \\ j=1,\ldots,n .
\label{eq.16}
\end{multline}
Thus no finite collection of bounded observables in the protocol algebra
\(C_S\) can operationally exclude the polymer representation on the basis of
expectation values alone.
The regularity of \(\pi\) is not used as an approximation assumption; it
specifies the usual Stone-von Neumann sector against which the non-regular
polymer representation is being compared.

\paragraph{Proof.}
The group \(\Gamma_S\) is a finitely generated abelian group, considered as a
discrete group.
The restriction of the Weyl multiplier to \(\Gamma_S\) is
\(
        c(\gamma,\gamma')
        =
        e^{-\frac{i}{2}\sigma(\gamma,\gamma')}.
\)
By the universal property of twisted group \(C^*\)-algebras,
the protocol algebra \(C_S\) is
canonically isomorphic to
the twisted group \(C^*\)-algebra
\(
        C_S\simeq C^*(\Gamma_S,c),
\)
with product
\(
        u_\gamma u_{\gamma'}
        =
        c(\gamma,\gamma')u_{\gamma+\gamma'} .
\)
Since \(\Gamma_S\) is abelian, it is amenable. Hence the full and reduced
twisted group \(C^*\)-algebras coincide \cite{Hulanicki,Rieffel}.

We use the following standard finite-set form of Fell's approximation theorem. If \(\sigma\) is a faithful
representation of a \(C^*\)-algebra \(A\), then every state \(\omega\) on
\(A\) can be approximated, on any prescribed finite set of elements of \(A\)
and to any prescribed accuracy, by finite convex combinations of vector
states belonging to \(\sigma\) \cite{FellDoran,KadisonRingrose}. 
Equivalently, for every finite set \(\mathcal F\subset A\) and every
\(\varepsilon>0\), there exist unit vectors \(\psi_1,\ldots,\psi_r\) in the
representation space of \(\sigma\) and weights \(t_k\ge0\), with \(\sum_{k=1}^r t_k=1\), such that
\begin{equation}
        \left|
        \omega(A)
        -
        \sum_{k=1}^r t_k
        \langle\psi_k,\sigma(A)\psi_k\rangle
        \right|
        <\varepsilon,
        \quad A\in\mathcal F .
\end{equation}
In physical terms, this means that a faithful representation contains enough vector states to reproduce arbitrary states on any finite list of tested observables. 
The Weyl \(C^*\)-algebra associated with a non-degenerate finite-dimensional
symplectic vector space is simple \cite{BratteliRobinson,Dixmier,Rieffel}. 
Therefore the nonzero polymer representation
\(
        \pi_{\rm pol}:\mathcal W\longrightarrow B(\mathcal H_{\rm pol})
\)
is faithful, and its restriction to the subalgebra \(C_S\) is faithful. Now
define the state on \(C_S\)
\(
        \omega(A)
        =
        \operatorname{Tr}\big(\rho\,\pi(A)\big),
        \,\, A\in C_S .
\)
Applying Fell's approximation theorem to the faithful representation
\(\pi_{\rm pol}|_{C_S}\), with
\(
        \mathcal F=\{A_1,\ldots,A_n\},
\)
gives unit vectors \(\psi_1,\ldots,\psi_r\in\mathcal H_{\rm pol}\) and
weights \(t_k\ge0\), \(\sum_{k=1}^r t_k=1\), such that
\begin{equation}
        \left|
        \omega(A_j)
        -
        \sum_{k=1}^r t_k
        \langle\psi_k,\pi_{\rm pol}(A_j)\psi_k\rangle
        \right|
        <\frac{\eta}{2},
        \,\, j=1,\ldots,n .
\end{equation}

It remains to ensure that the approximating polymer state is supported on
finitely many invariant protocol sectors. Finitely supported vectors are
dense in \(\ell^2(\mathbb R_d)\). Therefore each \(\psi_k\) may be replaced
by a sufficiently close finitely supported unit vector. 
This replacement changes all vector-state expectation values of the
contractions \(A_j\) by a controlled amount: if
\(\|\psi\|=\|\phi\|=1\) and \(\|A_j\|\le1\), then
\begin{equation}
        \left|
        \langle\psi,A_j\psi\rangle
        -
        \langle\phi,A_j\phi\rangle
        \right|
        \le
        2\|\psi-\phi\|.
\end{equation}
Choosing the replacements close enough, the total additional error can be
made smaller than \(\eta/2\) for all \(j=1,\ldots,n\).

The union of the finite supports of these approximating vectors is contained
in finitely many cosets of the shift subgroup \(\Lambda_S\) defined in
Eq.~\eqref{eq:LambdaS}. Let \(x_1,\ldots,x_N\) be representatives of these
cosets. Using the invariant sectors \(\mathcal H_{x_\alpha,S}\) defined in
Eq.~\eqref{eq:polymer-sector}, the approximating vectors therefore lie in
the finite direct sum
\begin{equation}
        \mathcal H_S^{\rm pol}
        =
        \bigoplus_{\alpha=1}^N
        \mathcal H_{x_\alpha,S}.
\end{equation}
Each \(\mathcal H_{x_\alpha,S}\) is invariant under
\(\pi_{\rm pol}(C_S)\). 
Indeed, the Weyl generators with labels in \(\Gamma_S\) shift basis vectors
only by elements of \(\Lambda_S\), and this invariance extends from finite
algebraic combinations to the norm-closed algebra \(C_S\) because
\(\mathcal H_{x_\alpha,S}\) is closed.

Consequently, \(\mathcal H_S^{\rm pol}\) is an invariant finite direct sum
of protocol sectors. Finally set
\begin{equation}
        \rho_{\rm pol}
        =
        \sum_{k=1}^r t_k |\psi_k\rangle\langle\psi_k| .
\end{equation}
This is a density operator supported on \(\mathcal H_S^{\rm pol}\).
Combining the Fell approximation error with the finite-support replacement
error gives Eq.~\eqref{eq.16}. This proves the theorem.

\paragraph{Finite tomography and operational consequences.}
The theorem has a direct interpretation for continuous-variable tomography.
A finite homodyne or displacement-tomography protocol uses finitely many
local-oscillator phases, finitely many displacement settings, finitely many
coarse-graining choices, and finitely many samples. The resulting data form
a finite list of experimentally estimated bounded statistics.
When these statistics are expressed, or approximated within the experimental
tolerance,
by a finite family \(A_1,\ldots,A_n\in C_S\), the theorem implies that their
expectation values in a regular representation can be reproduced 
by a suitable polymer state within the same prescribed tolerance.
Such a protocol may test many features of a state, a
gate, or measurement model, but it cannot certify the strong-continuity
condition required by the Stone-von Neumann theorem.

This also addresses the apparent operational concern raised by the existence of the polymer approximation. The result does not say that regular and polymer quantizations are physically identical in every respect. It proves the more precise statement that finite Weyl-displacement data alone cannot decide regularity. If additional physical structure is imposed, such as a strongly continuous time-evolution group, a Hamiltonian with appropriate domain properties,
energy regularity, semiclassicality conditions, or a continuum limit, then regular and polymer representations may become physically distinguishable.  Such distinctions arise from extra dynamical or analytic input, not from finite access to the Weyl relations alone.


The relation with polymer quantization and loop quantum gravity is structural rather than merely analogical. In the standard background-independent loop representation, polymer-type kinematics arise because the elementary variables are not the infinitesimal canonical fields themselves, but their exponentiated and smeared counterparts: holonomies of the Ashtekar-Barbero connection and fluxes of the densitized triad. The Ashtekar-Lewandowski representation therefore does not realize the connection as a weakly continuous operator-valued field in the same sense in which the Schr\"{o}dinger representation realizes canonical position and momentum through strongly continuous Weyl translations. Instead, it realizes finite holonomy transformations and flux operators on a non-regular kinematical Hilbert space, and the usual Stone-von Neumann uniqueness mechanism is replaced by uniqueness theorems adapted to diffeomorphism-covariant holonomy-flux algebras \cite{RovelliSmolin,AshtekarLewandowski,LOST}. The polymer representation studied here is the finite-dimensional prototype of this same mechanism: finite exponentiated translations are represented exactly, while the corresponding infinitesimal generators need not exist because weak or strong continuity fails \cite{AshtekarFairhurstWillis,Halvorson,CorichiVukasinacZapata}.

This observation motivates the precise question addressed in the present work. If polymer kinematics are not an arbitrary deformation but the natural kinematical language of loop quantized geometry, then one may ask whether analogous non-regular representations are operationally excluded for matter degrees of freedom. Standard quantum mechanics assumes the regular Schr\"{o}dinger representation, in which Weyl translations are strongly continuous and Stone's theorem reconstructs the usual self-adjoint canonical operators. Polymer quantization keeps the Weyl relations but drops this continuity requirement. The issue is therefore not whether the algebraic commutation relations are modified, but whether finite experimental access to Weyl observables can force the regularity assumption that selects the Schr\"{o}dinger sector. Our theorem answers this question in a restricted but sharp sense: for every regular Weyl representation, every normal state, and every finite family of bounded observables in a finitely generated protocol algebra, there exists a polymer state reproducing the same expectation values to arbitrary prescribed accuracy. Thus finite kinematical Weyl data alone do not operationally exclude polymer quantization for matter fields.

The result should not be read as claiming that matter polymer quantization is already physically established, or that polymer and Schr\"{o}dinger dynamics are phenomenologically equivalent. It says something more precise. The distinction between regular and polymer quantization cannot be decided by finite kinematical tests of the Weyl relations alone. If matter fields were fundamentally polymer quantized, ordinary Schr\"{o}dinger quantum mechanics would have to appear as an effective description selected by additional structure: Hamiltonian dynamics, semiclassical states, energy regularity, coarse graining, continuum reconstruction, or a low-energy limit. Conversely, possible deviations from standard quantum mechanics would not be expected merely from the algebraic Weyl relations, which both representations satisfy, but from the non-regular realization of dynamics and limits. In this sense, the present work clarifies the operational relation between standard quantum mechanics and the polymer kinematics underlying loop quantum gravity: regularity is not an experimentally forced consequence of finite Weyl data, but an additional analytic and physical assumption whose justification must come from dynamics and semiclassical recovery.

\paragraph{Conclusion.}
We have shown that the strong-continuity hypothesis entering the Stone-von Neumann uniqueness theorem cannot be certified by any fixed finite-resolution protocol probing Weyl translations. A finite protocol accesses only the norm-closed algebra \(C_S\) generated by a finitely generated subgroup of displacement labels. On this protocol algebra, expectation values produced by a regular Weyl representation can be reproduced, to arbitrary prescribed accuracy and for any finite family of bounded observables, by suitable polymer states. The polymer representation therefore supplies a concrete non-regular extension satisfying the Weyl relations exactly while failing the continuity assumption required for Stone-von Neumann uniqueness.  This conclusion leaves the Stone-von Neumann theorem intact. The theorem remains the correct uniqueness statement for irreducible regular representations. What our analysis clarifies is the operational status of regularity, it is an additional analytic and physical assumption, not something enforced by finite Weyl-displacement data. Finite agreement with Schr\"{o}dinger quantum mechanics therefore does not, by itself, exclude non-regular representations. To exclude or select among such representations one must impose further structure, for example Hamiltonian dynamics, energy regularity, semiclassicality, or a continuum limit. This is precisely the point at which physics beyond the finite Weyl protocol enters. 
In this restricted but important sense, polymer kinematics in quantum-gravity motivated quantizations are not ruled out by finite Weyl data alone, their physical viability is instead a question of additional dynamical and semiclassical input.

\paragraph{Conflict of Interest}
The authors declare that there are no conflicts of interest regarding this work.

\paragraph{Data Availability Statement}
This study contains no experimental data. All theoretical results are included in the manuscript.

\bibliographystyle{apsrev4-2}
\bibliography{q_unique}  

\end{document}